# Can multivariate Granger causality detect directed connectivity of a multistable and dynamic biological decision network model?

Abdoreza Asadpour[1] and KongFatt Wong-Lin[1]

[1] Intelligent Systems Research Centre, School of Computing, Engineering and Intelligent Systems, Ulster University, Magee campus, Derry~Londonderry, Northern Ireland, UK

**Abstract.** Extracting causal connections can advance interpretable AI and machine learning. Granger causality (GC) is a robust statistical method for estimating directed influences (DC) between signals. While GC has been widely applied to analysing neuronal signals in biological neural networks and other domains, its application to complex, nonlinear, and multistable neural networks is less explored. In this study, we applied time-domain multivariate Granger causality (MVGC) to the time series neural activity of all nodes in a trained multistable biologically based decision neural network model with real-time decision uncertainty monitoring. Our analysis demonstrated that challenging two-choice decisions, where input signals could be closely matched, and the appropriate application of fine-grained sliding time windows, could readily reveal the original model's DC. Furthermore, the identified DC varied based on whether the network had correct or error decisions. Integrating the identified DC from different decision outcomes recovered most of the original model's architecture, despite some spurious and missing connectivity. This approach could be used as an initial exploration to enhance the interpretability and transparency of dynamic multistable and nonlinear biological or AI systems by revealing causal connections throughout different phases of neural network dynamics and outcomes.

## 1 Introduction

The rapid advancement of modern artificial intelligence (AI) and machine learning has been significantly driven by the development and deployment of neural network models [1]. These models, particularly those exhibiting nonlinear dynamics, have shown remarkable capabilities in various complex tasks, including decision-making, pattern recognition, and predictive modelling [1]. Nonlinear neural networks, such as recurrent neural networks (RNNs), long short-term memory networks (LSTMs), and convolutional neural networks (CNNs) in deep learning architectures, are characterised by their ability to capture complex, time-dependent patterns and interactions within data [1].

Explainable AI (XAI) has emerged as a critical area of research, aiming to make the inner workings of dynamic AI models more comprehensible to humans. By elucidating how neural network nodes activate and interact over time, XAI can help in identifying the underlying mechanisms that drive model predictions and decisions [2]. Various methods have been proposed to achieve this, including feature attribution (e.g., SHAP values), model distillation, and visualisation techniques [3]. However, these approaches often fall short in providing a detailed understanding of the causal relationships between network components, especially in nonlinear settings [4] which occur frequently in biological systems [5]. For instance, feature attribution methods can indicate which inputs are important [3] but may not reveal how interactions between inputs evolve over time or contribute to the overall decision-making process. This is especially the case for multistable neural networks that can transition between multiple stable states based on input stimuli and noise perturbation [6].

Granger causality (GC), a statistical method for determining whether one time series can predict another, offers a promising approach to uncovering these causal connections. By estimating directed connectivity (DC), GC can reveal the directed influence between different nodes in a neural network, shedding light on the temporal dynamics of neural activation [7]. While GC has been widely applied to analysing neuronal signals in biological neural networks [8] and other domains [e.g., 9], its application to complex, nonlinear, and multistable neural networks, particularly in the context of decision uncertainty, is less explored. To test GC-DC analyses, linear autoregressive models are often used as ground truth [10]. Moreover, the models that generate the data for testing are usually abstract dynamical (especially autoregressive) models. In particular, there is no study that has tested, as ground truth, more realistic nonlinear, dynamic biologically based network model endowed with multistable states. This work will address this by testing multivariate GC (MVGC) on our previous stochastic, nonlinear dynamic mean-field model of decision-making, with real-time decision uncertainty monitoring and motor output corresponding to the choice made [11].

## 2 Methods

### 2.1 Model Description

We employed a multistable neural network model designed to simulate decision-making processes and real-time uncertainty monitoring [11]. The model consists of several interconnected neural units representing sensorimotor (SM) regions, inhibitory (INH) control, uncertainty encoding (U), and motor movement (Fig. 1A). Each neural unit can transition between multiple stable states based on input stimuli, reflecting the nonlinear dynamics inherent in complex decision-making tasks [11, 12]. This type of neural network is crucial for capturing the intricate temporal patterns and decision mechanisms observed in human and animal behaviour [11, 12] and, under certain tasks, can perform better than state-of-the-art deep networks systems [13].

## 2.2 Model Simulation

We simulated our previously developed nonlinear mean-field model for two-choice decision-making with real-time decision uncertainty monitoring and motor movement (to overtly report the made decision). The network includes different neural units: leftward (L-SM) and rightward (R-SM) receiving leftward stimulus input (L-IN) as well as rightward input (R-IN) for choice encoding in the SM region, INH control, uncertainty-encoding, as well as leftward (L-M) and rightward (R-M) for motor movement (Fig. 1A) [11]. The left/rightward stimulus definition is for the sake of convenience and can be generalised to other binary stimuli.

To test the model, we ran 2000 simulation trials for each level of sensory evidence quality ($\varepsilon$), which equates to the normalised difference in stimulus inputs to L-SM and R-SM (Fig. 1A). Sensory evidence quality represents how clear or strong the difference of the input signals is, similar to the concept of signal-to-noise ratio in signal processing. We explored a range of $\varepsilon$ values from 0% (very difficult task) to 100% (very easy task) and further verified our results with an additional 10,000 trials at lower qualities (0%, 3.2%, 6.4%, and 12.8%). These specific $\varepsilon$ values were chosen to systematically examine the network's performance across more difficult tasks ensuring that we could analyse the dynamics of the network under more uncertainty [11]. Due to the model's left-right symmetry, without loss of generality, rightward choice was defined as the correct response with generally larger stimulus input.

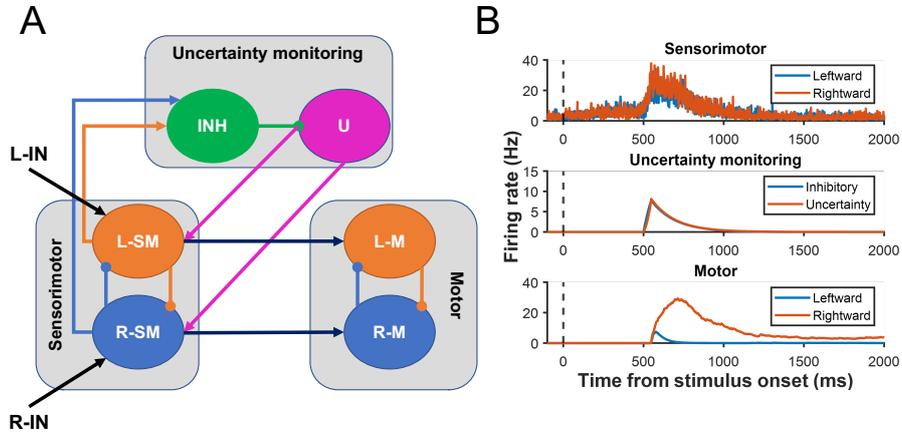

**Fig. 1.** (A) Original network model architecture [11]. (B) Sample model activities ($\varepsilon$=0%) for a single decision made (in this case, a right choice). See text for notations.

We grouped trials into correct and error choices, i.e. different network outcomes. A correct (error) choice is defined when the SM unit that first crosses some prescribed activity threshold (35.5 Hz) is in the same (opposite) direction as the direction of the overall signal, e.g. a rightward motor movement with net rightward signal. Then, we extracted stimulus- and decision-locked neural activities over time epochs (in time series format) for each $\varepsilon$ level. Stimulus- or response-locked activities were defined as aligning the activities at stimulus onset or decision time. respectively. A simulation trial generally began with the SM units competing in a winner-take-all (WTA) manner, before the INH and U units were transiently activated (once SM reached the decision threshold), followed by WTA behaviour of the motor movement neural units (Fig. 1B) [11]. Data were smoothed using a moving average filter with a duration of 50 ms and a stepsize of 5 ms. Inactive neural units were not considered. See [11] for further details and source codes for simulations.

### 2.3 Multivariate Granger Causality Analysis

Then the generated data were input into an MVGC toolbox [7]. Granger causality determines if one time series can predict another. If $X$ Granger-causes $Y$, past values of $X$ should help predict $Y$ beyond the past values of $Y$ alone [7]. Mathematically, $X$ Granger-causes $Y$ if:

$$\sigma^2_{Y|Y_{past}} > \sigma^2_{Y|X_{past},Y_{past}} \tag{1}$$

where $\sigma^2_{Y|Y_{past}}$ is the variance of the prediction error of $Y$ using its past values, and $\sigma^2_{Y|X_{past},Y_{past}}$ is the variance using past values of both $X$ and $Y$. The Vector Autoregressive (VAR) model is typically used:

$$Y_t = \sum_{i=1}^{p} a_i Y_{t-i} + \epsilon_t \tag{2}$$

$$Y_t = \sum_{i=1}^{p} b_i Y_{t-i} + \sum_{j=1}^{q} c_j X_{t-j} + \eta_t \tag{3}$$

Time-domain pairwise conditional GC (PWGC) extends GC to conditional relationships, measuring the influence of $X$ on $Y$ while controlling for $Z$. Models used are:

$$\text{Full model (with } X\text{): } Y_t = \sum_{i=1}^{p} c_i Y_{t-i} + \sum_{j=1}^{q} d_j Z_{t-j} + \sum_{k=1}^{r} e_k X_{t-k} + \epsilon_t \tag{4}$$

$$\text{Reduced model (without } X\text{): } Y_t = \sum_{i=1}^{p} a_i Y_{t-i} + \sum_{j=1}^{q} b_j Z_{t-j} + \epsilon'_t \tag{5}$$

where $\epsilon_t$, $\epsilon'_t$, and $\eta_t$ are error terms, and $p$, $q$ and $r$ are the maximum lags considered. G-causality from $X$ to $Y$ is the following log-likelihood ratio:

$$F_{X \to Y|Z} \equiv \ln\left(\frac{|\Sigma'_{YY}|}{|\Sigma_{YY}|}\right) \tag{6}$$

where $\Sigma'_{YY} = \text{cov}(\epsilon'_t)$ and $\Sigma_{YY} = \text{cov}(\epsilon_t)$ are residual covariance matrices for the reduced and full models [7].

To conform to quasi-stationary conditions for MVGC [7], we used sliding time window durations ranging from 100 to 400 ms, with a 50 ms step and a 90% overlap in addition to a window duration of 75 ms. The selection of these parameters was based on preliminary experiments and existing literature [14], which suggest that these window lengths are sufficient to capture dynamic interactions while maintaining stationarity. Shorter windows (less than 75 ms) were found to be inadequate as they did not create a positive-definite matrix necessary for calculating the autoregressive model. Stimulus-locked data began from 450 ms before the stimulus onset time to the fastest decision time across all trials for each evidence quality. For INH, U, and motor movement units, stimulus-locked data with decision time longer than 500 ms was used. Decision-locked data were similarly analysed using the fastest decision time.

PWGC for 2000 trials (a larger number of trials did not affect the results) and a significance level of 0.05 using an F-test with false discovery rate correction [15] was applied. The model order in MVGC was calculated using the Akaike information criterion (AIC), which is widely used for model selection due to its balance between goodness-of-fit and model complexity [7]. Self-connectivity within a neural unit was not considered since we utilised PWGC. MATLAB R2021b with the Northern Ireland High-Performance Computing (NI-HPC) facility and Kelvin2 system (www.ni-hpc.ac.uk) were used.

## 3   Results

For correct and error choices, stimulus-locked activities with low $\varepsilon$ (<4%) demonstrated significant PWGC between INH and U units with response times longer than 500 ms and a 100 ms sliding time window. For correct choices, these connections had less than 0.5 PWGC, but for error choices, they were robust ($\approx$1.9 PWGC) (Fig. 2). This was expected, as higher decision uncertainty was prone to more error choices [11]. Furthermore, INH and U correctly exhibited forward connections to both SM units. For correct choices, connections to L-SM were stronger, as compared to error choices, in which connections to R-SM were stronger (Fig. 2A). This indicated that evaluating both decision outcomes is needed. In stimulus-locked trials, no additional significant relationships have been observed perhaps due to strong bias of decision outcome and reduced connectivity. Higher $\varepsilon$ and longer sliding time windows did not reveal these observed connections. Some spurious connections were also found (compare Fig. 2 to Fig. 1).

The PWGC analysis of decision-locked data with low sensory evidence quality ($\varepsilon$ < 4%) and a 100 ms sliding window length covering 14 ms (the fastest) post-response onset revealed several key connections: (i) between INH and U units; (ii) between L-M and R-M units; (iii) from INH and U to both L-M and R-M units; (iv) from U to L-SM unit; (v) from L-SM to L-M unit; and (vi) from R-SM to U unit (Fig. 3). For correct choices, additional connections were observed from R-SM to both L-SM and R-M units (Fig. 3A). For error choices, an additional connection from INH to L-SM was noted (Fig. 3B). These results highlight how decision-locked analysis can uncover dynamic relationships that are not evident in stimulus-locked data, providing deeper insights into the network's behaviour during the decision-making process. Finally, by unifying the identified connections, we recovered most of the original model's connections but also found spurious connections due to indirect influences and missing connections due to low neural activity (Fig. 4).

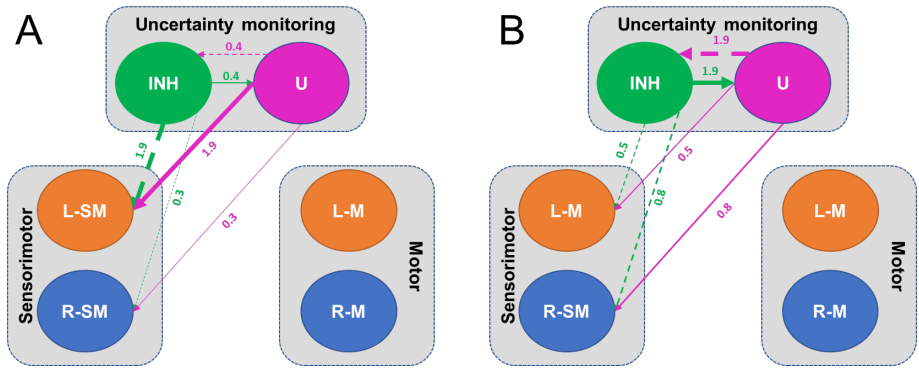

**Fig. 2.** MVGC connectivity with $\varepsilon < 4\%$ and 100 ms time window at 420-520 ms post-stimulus onset. Correct (A) and error (B) choices. Thicker connection: higher PWGC. Dashed: spurious. Numerics: PWGCs.

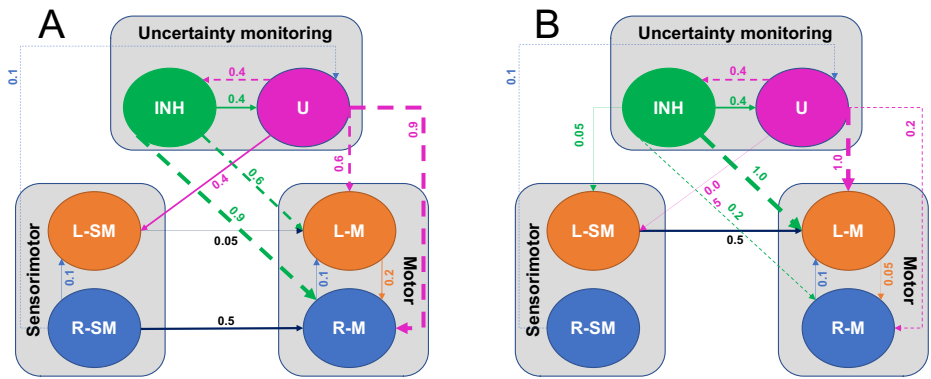

**Fig. 3.** MVGC connectivity for response-locked trials covering 14 ms post-response onset. Correct (A) and error (B) choices. Notations as in Fig. 2.

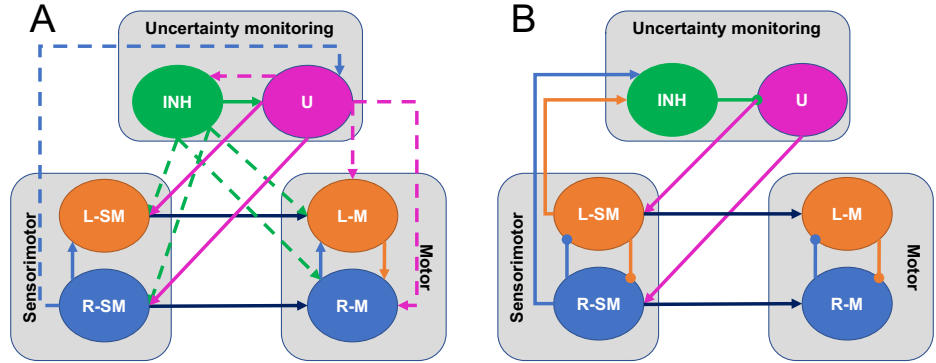

**Fig. 4.** Unified directed connections from Fig. 2 and Fig. 3 effectively recovered many of the connections from the original model. (A) Unified model. (B) Original model, for comparison.

## 4   Discussion

By applying multivariate Granger causality to a nonlinear neural network model exhibiting multistability, we were able to identify and quantify the directional influence between various neural units during the decision-making process. This approach complements existing XAI methods by uncovering the temporal dependencies and causal pathways that drive the network's behaviour. For instance, our analysis revealed how inhibitory and uncertainty-encoding units interact with sensorimotor and motor units under different decision conditions, highlighting the role of decision uncertainty in modulating network connectivity. Our results emphasise the necessity of applying difficult inputs to the network to thoroughly activate and recover many of the connections. Easy tasks may not sufficiently activate these connections, leading to an incomplete assaying of the network's dynamics. This also informs future cognitive neuroscience experiments to use difficult, more ambiguous stimuli. Further, future research could extend this analysis to contemporary deep and/or recurrent neural network architectures, exploring the utility of GC in elucidating their complex, hierarchical, and nonlinear connectivity patterns.

For nonlinear dynamic networks with temporal dynamics, it is also crucial to examine the interactions of connections for stimulus-locked (activities locked to stimulus input onset) and decision-locked (activities locked to decision onset). This approach is inspired by cognitive neuroscience [16]. This dual perspective provides a more comprehensive view of how different phases of neural activity contribute to the overall decision-making process. In addition, GC is computationally efficient, model-based, and offers superior sensitivity and specificity in detecting directed connectivity compared to other baseline approaches such as transfer entropy and partial directed coherence [17].

Despite the encouraging results, the current work has its limitations. Firstly, the model was not tested on standard deep network models that are inherently static [18]. However, there is a growing interest in dynamic neural networks in the AI/machine-learning community that can perform, for example, adaptive inference along the temporal dimension for sequential data such as videos and texts [18]. In particular, MVGC can act as an additional probe which allows for independent verification of the network's behaviour, ensuring that the changes observed are not influenced by internal network biases [19]. Additionally, this approach is beneficial when implementation of mechanisms to track and log changes internally is complex and might require significant modifications to the network's architecture [16]. Secondly, although our results could generally recover much of the information flow through the network model, they were not perfect as we found some spurious connections while some other direct connections were not detected. This could be due to the inherent standard measure used in the multivariate Granger causality [21]. That said, the spurious connections were mainly indirect connections – they were consistent with the combination of direct connections in the original models (Fig. 4). While this study relies on the MVGC toolbox, future work will explore other measures such as time-domain partial Granger causality, which may have better performance [22]. Further, we did not investigate and compare with other traditional XAI techniques [3]. Integrating these causal insights with traditional XAI techniques may perhaps further enhance the interpretability and transparency of neural network models. Lastly, it is important to note that GC relies on linearity and stationarity assumptions, which may not fully capture the complexities of nonlinear neural networks [7]. In our study, we mitigated these limitations by using fine-grained sliding time windows and verifying results with multiple trials to ensure robustness.

In summary, we have shown that multivariate Granger causality is promising in terms of uncovering nonlinear neural networks that have multistable states in the form of multiple decision options. This requires the consideration of appropriate time windows for the network's activity, the use of difficult task to explore the various possible network (decision) outcomes, and the unification of elucidated directed connectivity from different network outcomes. Such a set of approaches may potentially offer a useful exploratory tool to elucidate the connectivity of large and complex networks, leading to better interpretability.

## Acknowledgement

A.A. and K.W.-L. were supported by HSC R&D (STL/5540/19) and MRC (MC_OC_20020). We are grateful for access to the Tier 2 High-Performance Computing resources provided by the Northern Ireland High-Performance Computing (NI-HPC) facility funded by the UK Engineering and Physical Sciences Research Council (EPSRC), Grant No. EP/T022175/1.